\documentclass[aps,prl,prabib,twocolumn,showpacs,nofootinbib]{revtex4}
\usepackage{graphicx} \usepackage{amsmath} \usepackage{amssymb}
\usepackage{amsfonts} \usepackage{bm}

\begin{document}

\newcommand{\be}{\begin{equation}} \newcommand{\ee}{\end{equation}}
\newcommand{\bea}{\begin{eqnarray}}\newcommand{\eea}{\end{eqnarray}}

\title{Observations on spacetime symmetry and non-commutativity}

\author{Pulak Ranjan Giri} \email{pulakranjan.giri@saha.ac.in}

\affiliation{Theory Division, Saha Institute of Nuclear Physics,
1/AF Bidhannagar, Calcutta 700064, India}

\author{T. Shreecharan} \email{shreet@imsc.res.in}

\affiliation{The Institute of Mathematical Science, CIT Campus,
Taramani, Chennai 600113, India}

\begin{abstract}
We consider both the co-ordinates and momenta to be non-commutative
and define a non-commutative version of Lorentz symmetry which has a
smooth limit to the standard Lorentz symmetry. The
Poincar$\acute{e}$ algebra in this spacetime has also been
discussed.
\end{abstract}

\pacs{03.65.-w, 03.65.Db, 03.65.Ta}

\date{\today}

\maketitle

The study of non-commutative geometry and the formulation of field
theories in  such a space-time have become the subject of intense
investigation recently \cite{michael,rabin,saha,kumar,nair,bal1}. One
possible motivation is the  the appearance of the non-commutative
field theory as the limit of certain string theory  \cite{seiberg}.
However, non-commutativity is the intrinsic nature of quantum
theory. For example, the conjugate co-ordinates, $(x,p)$, of a
quantum mechanical model are non-commutative, leading to the famous
Heisenberg uncertainty relation
\begin{eqnarray}
\Delta x\Delta p\sim \hbar\,.
\end{eqnarray}
In quantum field theory the field operator  and its corresponding
momentum operator does not commute. It is obvious that in the
presence of spacetime non-commutativity the spacetime co-ordinates 
can not be measured simultaneous, leading to  an uncertainty relation.
For the co-ordinates $x,y$ being non-commutative one gets an
uncertainly relation of the form
\begin{eqnarray}
\Delta x\Delta y\sim \theta\,.
\end{eqnarray}
Incorporating spacetime non-commutativity naively in  standard quantum
mechanics and quantum field theory is fraught with difficulties and subtleties that 
one needs to tackle.

The introduction of spacetime non-commutativity in general modifies
the different symmetries of a system and it may break the symmetries
also. So, one of the prime objective is to investigate the spacetime
symmetries in non-commutative spacetime. It is known that when the
spacetime is non-commutative, with the non-commutative parameter
being constant, then the Lorentz symmetry of a system breaks down.
However it is possible to restore the Lorentz symmetry if one
considers the non-commutative parameter $\theta_{\mu\nu}$ to
transform under Lorentz transformation accordingly. However, in
$1947$ Snyder showed that it is possible to restore Lorenz
invariance by a non-commutative spacetime of a definite form
\cite{snyder,snyder1}. Although, the Poincar$\acute{e}$ symmetry is
usually lost.

In order to handle  the problems, associated  with the
non-commutativity, different forms of non-commutativity has been
introduced  in literature. For example in \cite{rabin}, the
non-commutativity of spacetime deforms the spacetime symmetry
algebra of a system. It has been shown that in relativistic case the
Lorentz, Poincar$\acute{e}$ and Conformal algebra is modified. They
obtained a one parameter family of conformal generators which form a
closed algebra whose commutative limit is smooth. The corresponding
non-relativistic counterpart of the symmetry algebra for example
Galilean, Schr$\ddot{o}$dinger, non-relativistic conformal algebra
are also studied in non-commutative space which leads to a deformed
algebra. Another example of getting a consistent Lorentz and
Poincar$\acute{e}$ invariance in non-commutative spacetime is to
twist \cite{chaichinan1,chaichinan2} the co-product defined by
$\Delta_\theta = \mathcal{T}^{-1}\Delta\mathcal{T}$, where
$\mathcal{T}=
\exp(-\frac{i}{2}\theta^{\mu\nu}\partial_\mu\otimes\partial_\nu)$.
This is known as   Drinfel'd twist.

The above mentioned prescription for implementing spacetime
symmetries  involve only co-ordinate space non-commutativity.
However in classical mechanics the phase space coordinates are
treated in equal footing. One can consider any set of the phase
space variables as   the coordinates of the system and the remaining
half then becomes the conjugate momenta. Therefore, there is no
reason to consider only coordinate non-commutativity and leaving the
momenta commutative. In fact it is known that, in quantum mechanics,
the generalized momenta $P_i$ in the magnetic field, $B$, background
do not commute, which leads to an uncertainty relation,
\begin{eqnarray}
\Delta P_1\Delta P_2\sim B\,.
\end{eqnarray}
In this letter we consider both the coordinates and momenta to have
deformation. We point out that such a thing is not very exotic and
has been introduced and studied previously
\cite{nair,bertolami,pulak}. We discuss the Lorentz
transformation in a phase space where both the co-ordinates and
momenta are non-commutative within themselves. Note that in
\cite{calmet} the non-commutative Lorentz symmetry has be obtained
by considering only  non-commutative space. They shows that there
exists an invariant length in non-commutative space which remains
invariant under Lorentz transformation. It must be pointed out that, what we call as Lorentz transformation is that set of transformations that leave the commutators (\ref{al1}) invariant.

We stat our discussion of this letter by taking  both the
co-ordinates and momenta to be non-commutative as mentioned above.
But then the Heisenberg uncertainty principle will be modified in
general. So, let us consider the phase space algebra
$\mathcal{A}_{\theta,\hat{\theta}}$ due to non-commutativity to be
of the form
\begin{eqnarray}
\nonumber \left[\hat{x}_\mu,\hat{x}_\nu\right]= i2\theta_{\mu\nu},~
\left[\hat{p}_\mu,\hat{p}_\nu\right]=i2\hat{\theta}_{\mu\nu}\,,\\
\left[\hat{x}_\mu,\hat{p}_\nu\right]= i\hbar(\delta_{\mu\nu}
+\theta_{\mu}^\beta\hat{\theta}_{\nu\beta})\,. \label{al1}
\end{eqnarray}
Here $\theta_{\mu\nu}$ and $\hat{\theta}_{\mu\nu}$ are
the non-commutativity parameters for co-ordinates and momenta
respectively. Note that the commutator between co-ordinates and its
conjugate momenta has been modified, which usually does not
happen if one considers only co-ordinates as non-commuting. One can
define a modified Planck's constant \cite{bertolami} like
$\hbar(\delta_{\mu\nu} +\theta_{\mu}^\beta\hat{\theta}_{\nu\beta})$.
This algebra, $\mathcal{A}_{\theta,\hat{\theta}}$, is not consistent
with the standard Lorentz symmetry because, the non-commutativity parameters
$\theta,\hat{\theta}$ are considered to be constant and therefore do
not transform under the Lorentz transformation. On the other hand
the Lorentz algebra is consistent with a phase space where both the
co-ordinate space and momentum space are commutative. The algebra
$\mathcal{A}$ being
\begin{eqnarray}
\nonumber \left[x_\mu,x_\nu\right]= 0,~
\left[p_\mu,p_\nu\right]=0\,,~ \left[x_\mu,p_\nu\right]=
i\hbar\delta_{\mu\nu}\,. \label{al2}
\end{eqnarray}
Note however that the commutative limit is smooth
\begin{eqnarray}
\lim_{\theta,\hat{\theta}\to
0}\mathcal{A}_{\theta,\hat{\theta}}=\mathcal{A}\,. \label{limit}
\end{eqnarray}
This can be taken as a guideline to get a non-commutative version of
Lorentz transformation which would be consistent with the algebra
$\mathcal{A}_{\theta,\hat{\theta}}$. In commutative space we know
that Lorentz transformation is defined as
\begin{eqnarray}
x^\prime_\mu= \Lambda_\mu^{~\nu} x_\nu,\quad p^\prime_\mu=
\Lambda_\mu^{~\nu} p_\nu\,,  \label{Lorentz}
\end{eqnarray}
which leave the lengths invariant
\begin{eqnarray}
S_x^2= x^\prime_\mu ~ x^{\prime \, \mu}= x_\mu x^\mu,~
S_p^2= p^\prime_\mu p^{\prime \, \mu} = p_\mu p^\mu\,.
\label{Lorentz1}
\end{eqnarray}
Since we want to know how the non-commutative phase space
co-ordinates, $\hat{x}_\mu,\hat{p}_\mu$, will transform under the
Lorentz transformation (\ref{Lorentz}) we need to know how they are
related to the commutative phase space co-ordinates $x_\mu,p_\mu$.
One possible representation is of the form
\begin{eqnarray}
\hat{x}_\mu=x_\mu -\theta_{\mu\nu} p^\nu,~~ \hat{p}_\mu=p_\mu
+\hat{\theta}_{\mu\nu} x^\nu\,,
 \label{rep}
\end{eqnarray}
It is equivalent to the relation
\begin{eqnarray}
x_\mu=\hat{x}_\mu + \theta_{\mu\nu} \hat{p}^\nu,~~ p_\mu=\hat{p}_\mu
-\hat{\theta}_{\mu\nu} \hat{x}^\nu\,, \label{rep1}
\end{eqnarray}
Let us now suppose that under Lorentz transformation the
non-commutative phase space co-ordinates transform as
\begin{eqnarray}
\hat{x}^\prime_\mu = x^\prime_\mu -\theta_{\mu\nu}
p^{\prime \, \nu},~~ \hat{p}^\prime_\mu = p^\prime_\mu
+\hat{\theta}_{\mu\nu} x^{\prime \, \nu} \,, \label{rep2}
\end{eqnarray}
which together with (\ref{Lorentz}) and (\ref{rep1}) gives the
generalization of Lorentz transformation
\begin{eqnarray}
\nonumber \hat{x}^\prime_\mu &=&(\Lambda_\mu^\alpha
+\theta_{\mu\nu}\Lambda^\nu_\beta \hat{\theta}^{\beta\alpha})\hat{x}_\alpha
+(\Lambda^\alpha_\mu\theta_{\alpha\delta}-\theta_{\mu\nu}\Lambda^\nu_\delta)\hat{p}^\delta\,,\\
\hat{p}^\prime_\mu&=&(\Lambda_\mu^\alpha
+\hat{\theta}_{\mu\nu}\Lambda^\nu_\beta\theta^{\beta\alpha})\hat{p}_\alpha
-(\Lambda^\alpha_\mu\hat{\theta}_{\alpha\delta}-\hat{\theta}_{\mu\nu} \Lambda^\nu_\delta)\hat{x}^\delta\,.
\label{rep3}
\end{eqnarray}
One can of course put a restriction
\begin{eqnarray}
\Lambda^\alpha_\mu\theta_{\alpha\delta}=\theta_{\mu\nu}\Lambda^\nu_\delta,~
\Lambda^\alpha_\mu\hat{\theta}_{\alpha\delta}=\hat{\theta}_{\mu\nu} \Lambda^\nu_\delta\,,
\label{restriction}
\end{eqnarray}
on the transformation (\ref{rep3}), which then becomes,
\begin{eqnarray}
\nonumber \hat{x}^\prime_\mu &=& (\Lambda_\mu^\alpha
+\theta_{\mu\nu}\Lambda^\nu_\beta\hat{\theta}^{\beta\alpha})\hat{x}_\alpha\,,\\
\hat{p}^\prime_\mu &=& (\Lambda_\mu^\alpha
+\hat{\theta}_{\mu\nu} \Lambda^\nu_\beta\theta^{\beta\alpha})\hat{p}_\alpha\,.
\label{rep4}
\end{eqnarray}
The restriction (\ref{restriction}) is only satisfied in
$2$-dimensions, but in more than $2$-dimensions it is not true in
general \cite{rabin}. Note that when both the non-commutativity
parameters are same $\theta=\hat{\theta}$ then the Lorentz
transformation for both co-ordinates and momenta are same in
$2$-dimensions. One can then define a generalized version of Lorentz
transformation due to non-commutativity as
\begin{eqnarray}
\widetilde{\Lambda_\mu^\alpha}= \Lambda_\mu^\alpha
+\theta_{\mu\nu}\Lambda^\nu_\beta\theta^{\beta\alpha}\,.
\label{glorentz}
\end{eqnarray}
We can also consider the Poincar$\acute{e}$ algebra, $\mathcal{P}$,
\begin{eqnarray}
\nonumber \left[J^{\alpha\beta}, J^{\gamma\delta}\right] &=&
i\left(-\eta^{\beta\gamma}J^{\alpha\delta}+\eta^{\alpha\gamma}J^{\beta\delta}
+\eta^{\delta\alpha}J^{\gamma\beta}-
\eta^{\delta\beta}J^{\gamma\alpha}\right),\\
\nonumber \left[p^{\alpha}, J^{\beta\gamma}\right]&=&
i\left(-\eta^{\alpha\beta}p^\gamma+
\eta^{\alpha\gamma}p^\beta\right),\\
\left[p^{\alpha}, p^{\beta}\right]&=& 0,\label{poincare}
\end{eqnarray}
where
\begin{eqnarray}
J^{\alpha\beta}= x^\alpha p^\beta- x^\beta p^\alpha\,. \label{jmu}
\end{eqnarray}
In the context of our non-commutative framework, the operator
$\hat{J}^{\alpha\beta}$, obtained replacing the phase space
co-ordinates $x_\mu, p_\mu$ by corresponding non-commutative counter
parts $\hat{x}_\mu, \hat{p}_\mu$ in  $J^{\alpha\beta}$,  becomes
\begin{eqnarray}
\nonumber \hat{J}^{\alpha\beta} &=& \hat{x}^\alpha \hat{p}^\beta -
\hat{x}^\beta \hat{p}^\alpha \\
\nonumber & = & x^\alpha p^\beta + x^\alpha\hat{\theta}^{\beta\nu} x_\nu
- \theta^{\alpha\mu}p_\mu p^\beta \\ &-& \theta^\alpha_\mu
p^\mu\hat{\theta}^{\beta\nu} x_\nu - <\alpha \leftrightarrows\beta
>\,, \label{jmunc}
\end{eqnarray}
where  $<\alpha \leftrightarrows\beta>$  refers to all previous
terms but with $\alpha$ and $\beta$ interchanged.
Note that the commutators among $\hat{J}^{\mu\nu}$ and
$\hat{p^{\mu}}$ will not close and therefore does not form any
algebra. However since we considered (\ref{rep3}) as our generalized
Lorentz transformation, which is consistent with the standard
Lorentz algebra in commutative space (\ref{Lorentz}), the
Poincar$\acute{e}$ algebra (\ref{poincare}) will also be the corresponding
algebra  in
non-commutative space. This is also commented in \cite{calmet}, but
there only co-ordinate space non-commutativity is considered.

In conclusion, we discussed the spacetime symmetry transformation
for example Lorentz and Poincar$\acute{e}$  symmetry in
non-commutative spacetime. We considered both the co-ordinates and
their conjugate to be non-commutative. The reason the spacetime
symmetry algebra remains invariant in non-commutative spacetime is
that the representation of non-commutative variables like
(\ref{rep}) is nothing but going to a new basis.

\section{Acknowledgment}
P. R. Giri thanks T. R. Govindarajan for giving him chance to attend
the workshop on ``Non-commutative geometry and quantum field theory"
held on 18 - 24 December, 2008 at IMSc, Chennai, India, where  part
of this work has been performed.

\end{document}